\documentclass[letterpaper,english,reprint,nofootinbib,aps,superscriptaddress,showpacs,showkeys,prl]{revtex4-1}

\usepackage{babel,calc,amsmath,amsthm,amssymb,graphicx,subfigure,xcolor,comment}
\usepackage{txfonts}
\usepackage{mathdots}
\usepackage{todonotes}
\usepackage[T1]{fontenc}
\usepackage[unicode=true]{hyperref}
\usepackage{booktabs}
\usepackage{threeparttable}
\usepackage{times}
\usepackage{CJK}
\hypersetup{
	colorlinks=true,       		
	linkcolor=blue,          	
	citecolor=blue,            
	urlcolor=black,           	
}

\theoremstyle{definition}

\def\bra#1{\left\langle {#1} \right\rvert}
\def\ket#1{\left\lvert {#1} \right\rangle}

\begin{document}

\begin{CJK*}{UTF8}{gbsn}

\title{Postselection-Free Cavity-Enhanced Narrow-Band Orbital Angular Momentum Entangled Photon Source}

\author{Pei Wan (万佩)}
\altaffiliation{These authors contributed equally to this work.}
\author{Wen-Zheng Zhu (朱文正)}
\altaffiliation{These authors contributed equally to this work.}
\author{Yan-Chao Lou (娄严超)}
\altaffiliation{These authors contributed equally to this work.}
\author{\\Zi-Mo Cheng (程子默)}
\author{Zhi-Cheng Ren (任志成)}
\affiliation{National Laboratory of Solid State Microstructures, School of Physics, Nanjing University, Nanjing 210093, China}
\affiliation{Collaborative Innovation Center of Advanced Microstructures, Nanjing University, Nanjing 210093, China}

\author{Han Zhang (张涵)}
\email[]{zhanghan@nju.edu.cn}
\affiliation{National Laboratory of Solid State Microstructures, School of Physics, Nanjing University, Nanjing 210093, China}
\affiliation{Collaborative Innovation Center of Advanced Microstructures, Nanjing University, Nanjing 210093, China}
\affiliation{Hefei National Laboratory, Hefei 230088, China}

\author{\\Xi-Lin Wang (汪喜林)}
\email[]{xilinwang@nju.edu.cn}
\affiliation{National Laboratory of Solid State Microstructures, School of Physics, Nanjing University, Nanjing 210093, China}
\affiliation{Collaborative Innovation Center of Advanced Microstructures, Nanjing University, Nanjing 210093, China}
\affiliation{Hefei National Laboratory, Hefei 230088, China}
\affiliation{Synergetic Innovation Center of Quantum Information and Quantum Physics, University of Science and Technology of China, Hefei 230026, China}

\author{Hui-Tian Wang (王慧田)}
\email[]{htwang@nju.edu.cn}
\affiliation{National Laboratory of Solid State Microstructures, School of Physics, Nanjing University, Nanjing 210093, China}
\affiliation{Collaborative Innovation Center of Advanced Microstructures, Nanjing University, Nanjing 210093, China}
\affiliation{Collaborative Innovation Center of Extreme Optics, Shanxi University, Taiyuan 030006, China}
\date{\today}

\begin{abstract}
Cavity-enhanced spontaneous parametric down-conversion (SPDC) provides a significant way to produce $\sim$10 MHz narrow-band photon pairs, which matches the bandwidth of photon for quantum memory. However, the output photon pairs from the cavity is not entangled and the postselection is required to create the entanglement outside the cavity, so the direct output of cavity-enhanced narrow-band entangled photon pairs is still an open challenge. Here we propose a solution that realizes the first postselection-free cavity-enhanced narrow-band entangled photon pairs. The entanglement is achieved in degree of freedom (DOF) of orbital angular momentum (OAM) by implementing an OAM-conservation SPDC process in an actively and precisely controlled cavity supporting degenerate high-order OAM modes. The measured linewidth for the two photons is 13.8 MHz and the measured fidelity is 0.969(3) for the directly generated OAM entangled two photons. We deterministically transfer the OAM entanglement to polarization one with almost no loss and obtain polarization entangled two photons with a fidelity of 0.948(2). Moreover, we produce narrow-band OAM-polarization hyperentangled photon pairs with a fidelity of 0.850(2) by establishing polarization entanglement with preservation of OAM entanglement, which is realized by interfering the two photons on a polarizing beam splitter (PBS) and post-selecting the events of one and only one photon in on each of the PBS port. Novel cavity may find applications in cavity-based light-matter interaction. Our results provide an efficient and promising approach to create narrow-band entangled photon sources for memory-based long-distance quantum communication and network. 
\end{abstract}
	
\maketitle

\end{CJK*}
Quantum memory~\cite{Chaneliere2005, Lvovsky2009} for storing the quantum state of single photons plays an essential role in realizing memory-based long-distance quantum communication~\cite{Duan2001, Sangouard2011}, large-scale quantum computers~\cite{Kok2007} and quantum networks~\cite{Wehner2018, Azuma2023}. For the quantum memory with atomic ensemble~\cite{Zhang2011, Ding2013, Nicolas2014, Parigi2015, Vernaz-Gris2018, Wang2019, Ye2022, Dong2023} and solid state system~\cite{Rakonjac2021, Lago-Rivera2021, Rakonjac2022}, the typical bandwidth of the memory is in the order of several MHz, which is 6 orders of magnitude less than the bandwidth of the photons generated by spontaneous parametric down-conversion (SPDC) in free space restricted by phase matching (typically in order of terahertz). It seems that passive filtering using etalons can solve this issue, but photon brightness will be dramatically reduced to a very low level and signal-to-noise ratio will inevitably decrease greatly. A good solution is cavity-enhanced SPDC~\cite{Ou1999, Scholz2009-OC, Slattery2019}, where a nonlinear crystal is inserted into the cavity as an active filter to compress photon bandwidth via cavity modes. The cavity-enhanced SPDC will greatly enhance the generation probability of down-converted photons in the cavity resonance mode with respect to the passively filtering scheme. Another solution to produce narrow-band photon source is spontaneous four-wave mixing (SFWM) in an atomic medium including cold atomic ensembles ~\cite{Kuzmich2003, Thompson2006, Yan2011, Srivathsan2013, Liao2014, Chen2015, Lee2016, Zhao2019, Mei2020} and hot atom vapor~\cite{Shu2016, Zhu2017, WangC2018, Hsu2021, Chen2022, Ma2024}, which relies precise control of atomic medium with lasers. Usually the atoms need to be cooled down to obtain narrow bandwidth in the order of several MHz~\cite{Kuzmich2003, Thompson2006, Yan2011, Srivathsan2013, Liao2014, Chen2015, Lee2016, Zhao2019, Mei2020} and the bandwidth of photons from SFWM in hot atom vapor will be often 1-2 orders of magnitude larger. While, this challenge was overcome by a series of recent methods including the paraffin coating, all-copropagating configuration and so on, which enable the generation of subnatural-linewidth biphotons by SFWM in hot atom vapor~\cite{Shu2016, Zhu2017, WangC2018, Hsu2021, Chen2022, Ma2024} with much lower apparatus complexity than cold-atom based SFWM.

Besides the narrow-band requirement, for application of memory-based quantum information processing, the photons of quantum source should also be entangled. For the photon pairs produced by the cavity-enhanced SPDC, whether two photons are entangled depends on the resonance modes, which are mainly controlled by both spatial and polarization modes. The spatial modes depend on the types of modes that can be supported in the cavity, while previous cavity-enhanced SPDC are restricted to the Gaussian mode. Moreover, the polarization modes must satisfy the phase matching requirement in the SPDC. In type-I cavity-enhanced SPDC~\cite{Ou1999, Lu2003, Liu2020}, it supports only one polarization resonance mode because two down-converted photons have the same polarization. The first cavity-enhanced SPDC~\cite{Ou1999} was realized in the single resonance manner. However, for the two photons in just one single resonance mode, it is difficult to deterministically separate them in different propagating paths, and more difficult to prepare two entangled photons. 

To obtain two resonance modes, type-II SPDC can generate the down-converted photons with orthogonal polarizations. With a careful design, dual-resonance type-II cavity-enhanced SPDC~\cite{Kuklewicz2006, Bao2008, Scholz2009, Wolfgramm2011, Rambach2016, Moqanaki2019, Prakash2021} can output narrow-band photon pairs in orthogonal linear polarizations, which can be deterministically separated into two paths by a polarization beam splitter (PBS). Nondegenerate type-I cavity-enhanced SPDC in frequency (such as 606 and 1436 nm~\cite{Fekete2013, Lago-Rivera2021}) provides another method to support two resonance modes, thus two down-converted photons can be outputted in different paths from the cavity~\cite{Lago-Rivera2021} or be deterministically separated into two paths by a dichroic mirror (DM)~\cite{Fekete2013}. However, two resonance modes are only directly two-path-separated, rather than entangled photons. To prepare the entanglement, postselection~\cite{Bao2008, Rakonjac2021}, by selecting partial two-photon components under special conditions, has to be introduced, but there are two costs: (1) losing a half two-photon brightness and (2) prohibiting two-photon interference due to the inherent noise from the abandoned components during post-selection process. The postselection manner strongly restricts the wide applications of entangled photon source. Direct generation of narrow-band entangled photon pairs without post-selection has been a long-term open challenge for the cavity-enhanced SPDC. 

Here we propose a solution, which introduces a new degree of freedom (DOF)---orbital angular momentum (OAM), to realize postselection-free narrow-band entangled photons from an actively and precisely controlled stable cavity. The crucial step is to break the limitation of \textit{two} cavity resonance modes, which can only make it possible to deterministically divide the two photons into two different paths. To create the entanglement, at least \textit{four} cavity resonance modes are required: two of them for the path separation of the down-converted photons and the other two for the entanglement preparation. It should be noted that the experimental difficulty will dramatically increase as the resonant modes increases. Up to now, for single-longitudinal-mode narrow-band photon pairs generated in the cavity-enhanced SPDC, two modes can be supported in the cavity, but four modes are very difficult to resonate simultaneously. Therefore, how to support four cavity modes become the most crucial factor in creating entanglement. Here we solve this problem by using the two degenerate spatial high-order cavity modes (i.e. Laguerre-Gaussian (LG) modes with $(p, m) = (0, \pm 1)$, where $p$ denotes the radial mode and chosen as $p = 0$ in this work, and $m$ is the integer topological charge denoting the azimuthal mode). Since such a mode will carry OAM of $m\hbar$ per photon~\cite{Allen1992}, the mode is also known as OAM mode. The OAM modes with $m = \pm 1$ are degenerate and can be supported in the same cavity. So that we can construct a cavity that allows two degenerate $m=\pm1$ OAM modes and two polarization modes to form \textit{four} modes for entanglement preparation. In particular, two-photon OAM entanglement can be also produced in collinear type-II SPDC~\cite{Hendrych2012, Malik2016, Bavaresco2018, Ecker2019} due to the OAM conservation~\cite{Mair2001, Pires2010}. As a result, the dual-resonance cavity-enhanced type-II SPDC supporting high-order OAM modes is a feasible route to direct generate path-separated narrow-band entangled photon pairs.

\begin{figure}[!htb]
	\centering
	\includegraphics[width=0.95\linewidth]{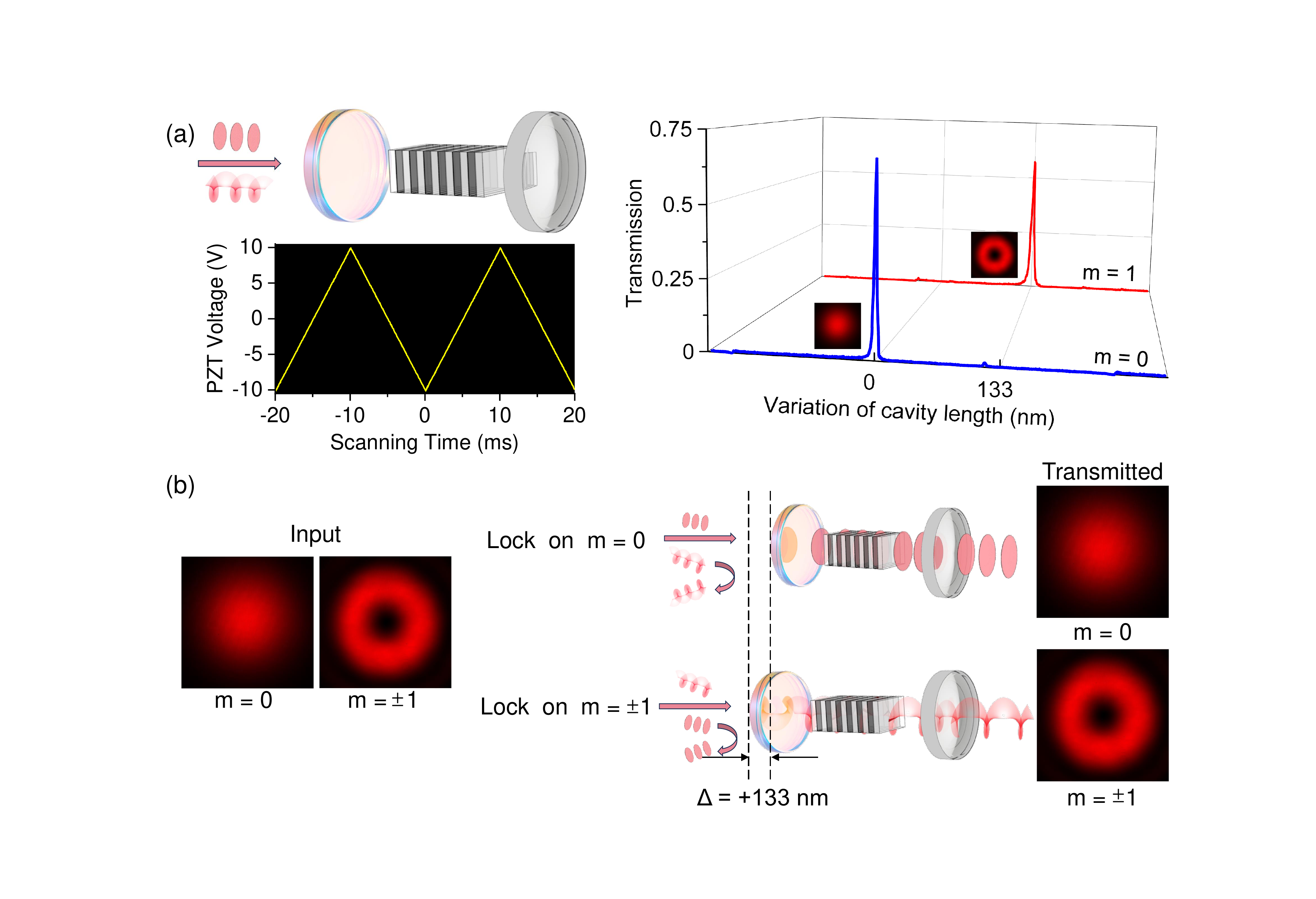}
	\caption{Principle to construct a high-order OAM cavity. (a) The structure of the FP cavity and the measured transmission for various input OAM modes with $m = 0$ and $m = \pm 1$. The cavity is composed of a PPKTP crystal, a concave mirror, and a flat mirror adhered to PZT that is used to scann the cavity length. The shift of the transmission peak shows clearly the dependence of the resonant OAM mode on the cavity length. (b) Locked cavity on the OAM modes of $m =0$ and $m = \pm 1$. Input OAM modes of $m =0$ and $m=\pm1$, only the matched mode can be transmitted.The laser transverse profiles, the transmission peaks curve and sweeping waveform are all experimental results.}
	\label{fig1}
\end{figure}

The LG modes are eigenstates in a cylindrical system, which provides the possibility to construct the cavity supporting high-order LG modes~\cite{Martinelli2004, Navarrete-Benlloch2008, Lassen2009, Santos2009, Liu2014, Cheng2018}. The crucial parameter in the cavity supporting OAM modes is Gouy phase~\cite{Zhou2017, Gu2018, Barros2021, Hiekkamlaki2022} as 
\begin{equation}
\Phi_{G}(z)=-(2p+|m|+1){\rm{arctan}}(z/z_{R}),
\label{eq1}
\end{equation}  
where $z$ is the propagation distance from the beam waist and $z_{R}$ is the Raleigh range of the Gaussian mode. Restricted by the self-consistent condition for the resonance mode in a cavity, the cavity length $L$ depends on the mode parameters of $(p, m)$, due to the existence of accumulated Gouy phase $\Phi_{G}(L)$ after one round trip in the cavity. Compared to the cavity supporting the Gaussian mode with $(p,m) = (0,0)$, we can increase the cavity length to support high-order LG modes~\cite{SuppMat}
\begin{equation}
\Delta L = (\lambda / 2\pi) (2p+|m|){\rm{arctan}}(L/z_{R}),  
\label{eq2}
\end{equation} 
where $\lambda$ is the wavelength of light. As shown in Eq.~(\ref{eq2}), the OAM modes with opposite topological charges are degenerate, such as two $m \! = \! \pm 1$ OAM modes in this work. The degenerate property for high-order OAM cavity meets the requirement for direct output of narrow-band entangled photon pairs. 

In experiment, we design and construct a cavity consisting of two mirrors and a 5-mm-long type-II  periodically poled KTiOPO$_4$ (PPKTP) crystal, as shown in Fig.~\ref{fig1}(a). The operating wavelength of our cavity is chosen to be 795 nm, corresponding to the 87Rb D1 line. Our cavity is a flat-concave cavity with an optical length of $L = 75$ mm, composed of a high-reflection coated flat mirror ($R>99\%$) controlled by a piezoelectric transducer (PZT) and a concave mirror ($R>97\%$). Therefore, we can precisely adjust the cavity length by the voltage applied on the PZT and obtain the dependence of transmission on the cavity length for various OAM modes as shown in Fig.~\ref{fig1}(a). The cavity length for the OAM modes with $m = \pm 1$ is 133 nm longer than the one with $m=0$. Then, by using the standard Pound-Drever-Hall (PDH) scheme~\cite{Black2001, Wang2015}, we can lock the cavity for various OAM modes with high quality as shown in Fig.~\ref{fig1}(b). 

Besides the two degenerate OAM modes with $m = \pm 1$, our cavity-enhanced SPDC requires the dual resonance for both horizontal ($H$) and vertical ($V$) polarization modes simultaneously. We firstly lock our cavity by using a frequency-locked horizontally linearly polarized laser ($\lambda=795$ nm corresponding to the the center frequency of $\omega_0$, bandwidth below 5 kHz) with OAM mode of $m=\pm1$ under the repetition rate of 40 Hz to eliminate the long-term drift noises, which are the main noises in our cavity system. The $H$ polarized mode will be well guaranteed for the resonance at $\omega_0$ with this active locking system. Furthermore, the $V$ polarized mode resonance at $\omega_0$ is simultaneously realized through carefully tuning the temperature of the PPKTP with the high precision of $0.003^\circ \rm{C} $.

Next we will investigate the quantum state generated by our cavity-enhanced SPDC. In two DOFs of OAM and polarization, four available computational states ($\ket{+1}_H$, $\ket{+1}_V$, $\ket{-1}_H$, $\ket{-1}_V$) for generated photons are determined by the designed dual-resonance cavity, while the combination and coherent superposition of these states are determined by the SPDC in the PPTKP pumped by the Gaussian mode with 0 OAM value. In this SPDC process, the OAM is conserved~\cite{Malik2016, Bavaresco2018, Ecker2019}, that is, the sum of the two down-converted photons' OAM values is equal to the OAM value of the pump photon. Under the pumping with 0 OAM value, the generated biphoton form our high order cavity should have a state of $\frac{1}{\sqrt2}(\ket{+1}_H\ket{-1}_V+\ket{-1}_H\ket{+1}_V)$. In the DOF of frequency, the ideal single longitudinal mode at center frequency of $\omega_0$ will be mixed with some nearby background modes~\cite{Bao2008} and the resulting output biphoton state should be
\begin{equation}
\begin{aligned}
\ket{\Psi}_C \propto & \left( \ket{+1}_H\ket{-1}_V+\ket{-1}_H\ket{+1}_V \right) \otimes \left[ \ket{\omega_{0}}_H\ket{\omega_{0}}_V \right.\\
& + {\sum}_{m=1}^{m=N}\sqrt{c_m}(\ket{\omega_{0}+m\Omega}_H\ket{\omega_{0}-m\Omega}_V \\
& \qquad \qquad \qquad \left. + \ket{\omega_{0}-m\Omega}_H\ket{\omega_{0}+m\Omega}_V) \right],
\label{eq3}
\end{aligned}
\end{equation}
where $c_m$ is the spectral ratio of the $m$th spectral line counting from the central frequency, $N$ is a half of the number of biphoton frequency modes within the phase-matching bandwidth of PPKTP, $\Omega=(\Omega_H+\Omega_V)/2$ is the average free spectral range (FSR) for the $H$ and $V$ polarized modes. In our experiment, $N=17$, $\Omega=1.8$~GHz and the difference between the FSRs of $H$ and $V$ modes is $\Delta \Omega=(\Omega_V-\Omega_H)=17$~MHz. The finesse of our cavity is $\sim$120. With $c_m$, we calculate the ratio of the summation of nearby background modes to the central frequency mode is $\sim$0.87 (see Fig.~S1 in~\cite{SuppMat}). To eliminate the nearby background modes, we use an etalon (FSR = 10.4~GHz, Finesse = 30) with the temperature control precision of $0.003^\circ$C. Finally, we obtain the output biphoton state with single longitudinal mode at the central frequency of $\omega_0$ in the form of
\begin{equation}
\ket{\Psi}_f=\frac{1}{\sqrt{2}}(\ket{+1}_{H}\ket{-1}_{V}+\ket{-1}_{H}\ket{+1}_{V}),
\label{eq4}
\end{equation}
which is easily separated to two paths by a PBS and generates the postselection-free narrow-band OAM-entangled photon pairs.

\begin{figure}[th]
	\centering
	\includegraphics[width=0.95\linewidth]{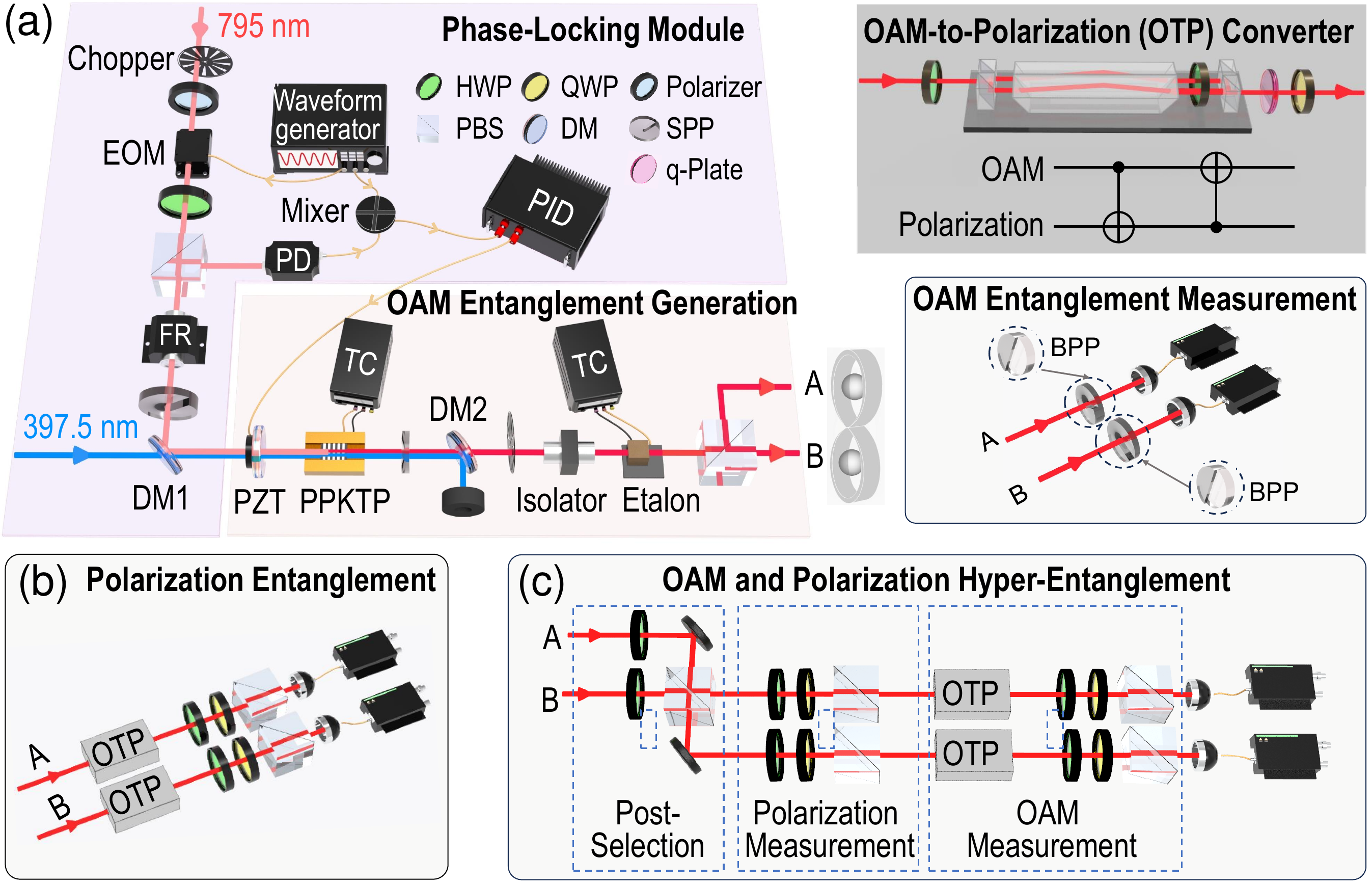}
	\caption{Direct generation of cavity-enhanced narrow-band OAM-entangled photon pairs. (a) Experimental setup for directly outputting the OAM-entangled paired two photons A and B from locked high-order degenerate OAM cavity-enhanced SPDC and the measurement scheme is composed of SPP or BBP and a single mode fiber. (b) Preparation and measurement scheme of polarization entanglement. The preparation is realized by two OAM-to-polarization converters in the right up of (a). (c) The OAM-polarization hyperentanglement generation by postselection on a PBS without destroying the original OAM entanglement. The OAM entanglement is achieved by the SPDC of OAM conservation under the pumping of 0 OAM value and the OAM conversion elements including q-plate, SPP, BPP, are utilized in preparation of locking laser and measurement of entangled biphoton state. EOM---elcctro-optic modulator, PD---power detector, PID---proportional-integral-derivative controller, FR---Faraday rotator, TC---temperature controller.}  
	\label{fig2}
\end{figure}
Figure~\ref{fig2} shows the experimental setup to directly generate narrow-band entangled photon pairs through the cavity-enhanced SPDC. The PPKTP in the locked cavity is pumped by a frequency-locked ultraviolet laser ($\lambda=397.5$ nm, bandwidth < 5 kHz), which is carefully aligned and focused to match the down-converted photons with the cavity mode~\cite{Boyd2020}. Two dichroic mirrors (DM1 and DM2) are utilized to spatially recombine the locking beam (reflected) and the pump beam (transmitted) in the front of the cavity, and spatially separate the pump beam (reflected) and the locking beam together with the down-converted photon pairs (transmitted) behind the cavity. The separation of the locking beam and the down-converted photon pairs is realized by the time division multiplexing with the duty ratio of 1:3 by a mechanical Chopper. In the following step, the down-converted photon pairs pass through the temperature-controlled etalon to eliminate the nearby background frequency modes from the cavity. There is a polarization-independent isolator between the cavity and the etalon to avoide the resonance between them. Next, the down-converted photon pairs are spatially separated by a PBS. The $V$ ($H$) polarized photon will propagate along the up (down) path and is denoted as A (B). Then the expected spatially separated narrow-band postselection-free OAM-entangled two photons is produced as
\begin{equation}
\ket{\Psi}=\frac{1}{\sqrt{2}}(\ket{+1}_{A}\ket{-1}_{B}+\ket{-1}_{A}\ket{+1}_{B}).
\label{eq5}
\end{equation}

\begin{figure}[!htb]
	\centering
	\includegraphics[width=0.95\linewidth]{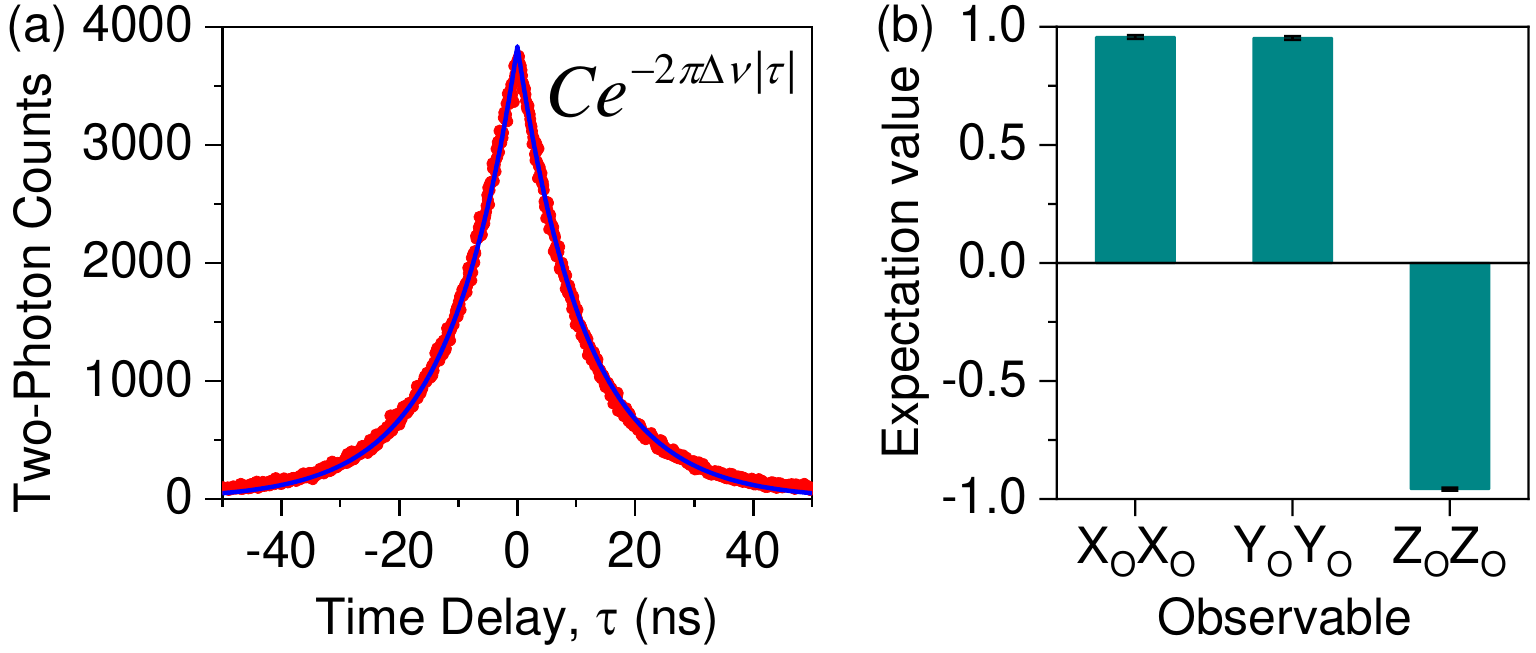}
	\caption{Time correlation measurement and experimental results of narrow-band OAM-entangled two photons. (a) Measured data and the fitting curve by a function of $C e^{-2\pi \Delta \nu |\tau|}$. The full width at half maximum (FWHM) of the fitting curve is $\sim$16.0 ns. (b) Measured expectation values of the joint observables $X_oX_o$, $Y_oY_o$ and $Z_oZ_o$.}
	\label{fig3}
\end{figure}

The verification of the generated photon pairs depends on the measurement of their two distinct features, namely narrow-band and OAM-entanglement. To verify the `narrow-band' feature, we measure the time correlation of the biphotons A and B (in the OAM basis of $\ket{+1}$and $\ket{-1}$) by recording their arrival times at the detectors with a time-to-digital-converter and obtain the time correlation curves as shown in Fig.~\ref{fig3}(a). Fitting with the theoretical predicted time correlation function~\cite{Scholz2009-OC, Slattery2019} in the form of $C e^{-2\pi \Delta \nu |\tau|}$, where $\Delta \nu$ is the linewidth and $\tau$ is the time delay between the two photons, we calculate the linewidth of our down-converted photons to be $\sim$13.8 MHz, which is well within the cavity linewidth~\cite{Slattery2019} and verifies the `narrow-band' feature for our generated photon pairs. To confirm the `OAM-entangled' feature, we can investigate the fidelity for the prepared two-photon states with respect to the ideal state in Eq.~(\ref{eq5}) through measuring the two photons in the eigenstate basis of $X_oX_o$, $Y_oY_o$ and $Z_oZ_o$ with sprial phase plates (SPPs) and binary phase plates (BPPs) together with single-mode fibers (SMFs)~\cite{Wang2015-N} as shown in the lower right corner of Fig.~\ref{fig3}(a), where $X$, $Y$ and $Z$ are the Pauli operators and the subscript `$o$' indicates the DOF of OAM. The density matrix $\rho_o=\ket{\Psi}\bra{\Psi}$ can be decomposed as $\rho_o = \frac{1}{4} (II + X_oX_o + Y_oY_o - Z_oZ_o)$, where $I$ is an identity matrix. Then we can calculate the fidelity with the expectation values of the joint observables of $X_oX_o$, $Y_oY_o$ and $Z_oZ_o$. With our experimental results in Fig.~\ref{fig3}(b), we obtain a fidelity of $0.969\pm0.003$ for our generated OAM-entangled state, which implies the high quality of our narrow-band entangled source.

\begin{figure}[!htb]
	\centering
	\includegraphics[width=0.95\linewidth]{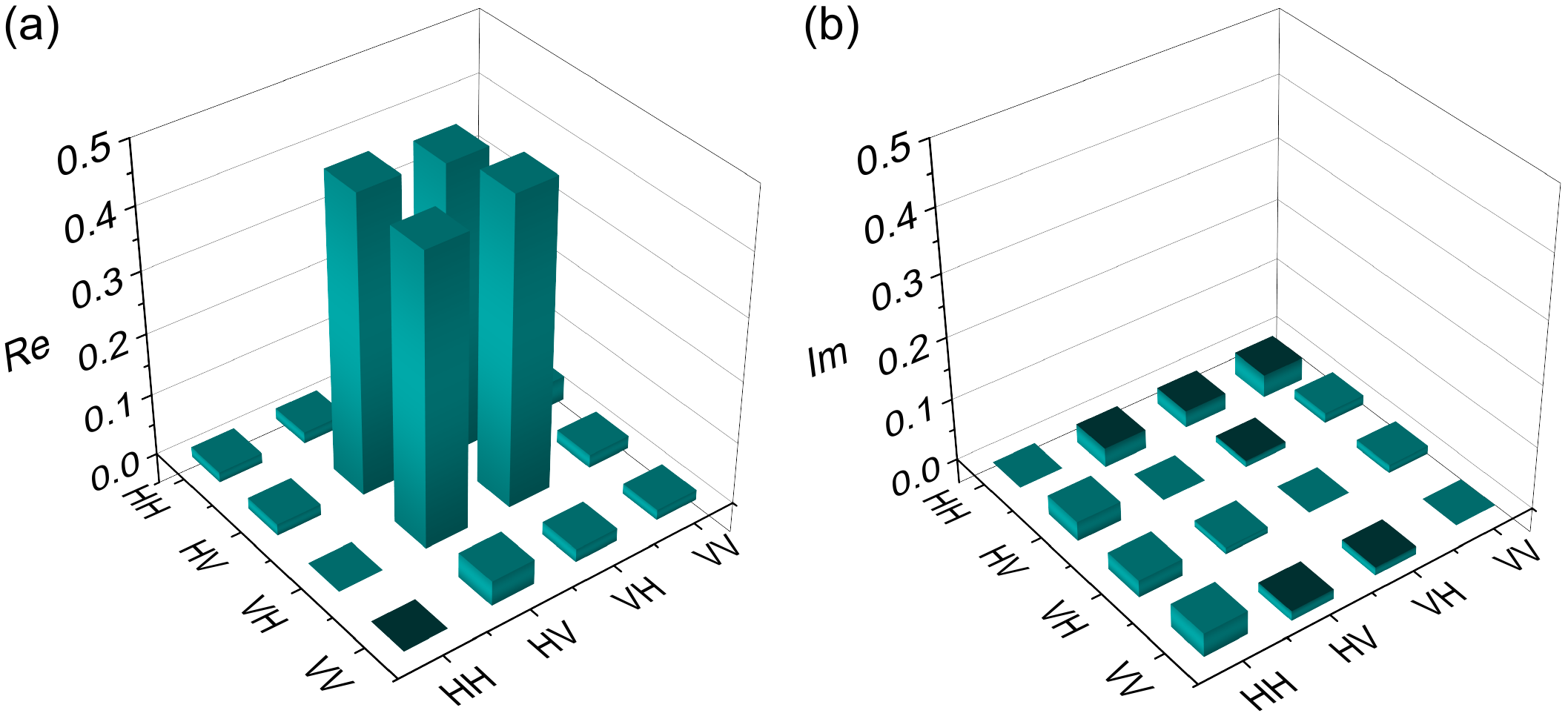}
	\caption{Reconstructed density matrix for the converted two-photon polarization entangled state. (a) Real part. (b) Imaginary part.}
	\label{fig4}
\end{figure}

Since the polarization is a widely used DOF in quantum information processing, one question is naturally raised that whether our system can produce narrow-band polarization-entangled source. Of course, the answer is completely affirmative. In fact, our photon entanglement in the  DOF of OAM can be deterministically transferred to one in the DOF of polarization by a OAM-to-polarization (OTP) converter~\cite{Wang2018, Xu2022}, as shown in the upper right corner of Fig.~\ref{fig3}(a), which is essentially a swap gate between OAM and polarization. As shown in Fig.~\ref{fig3}(b), using the two OTP converters to both A and B photons, we can transfer the original OAM entanglement to polarization entanglement with almost no loss and the polarization-entangled two photons should have the form of   
\begin{equation}
\ket{\Psi}_P=\frac{1}{\sqrt{2}}(\ket{H}_{A}\ket{V}_{B}+\ket{V}_{A}\ket{H}_{B}),
\label{eq6}
\end{equation}
which is one of four Bell states. To completely characterize the polarization entanglement, we perform a quantum state tomography~\cite{James2001, Lou2024} for our narrow-band polarization-entangled photon pairs. As the experimental results of tomography shown in Fig.~\ref{fig4}, we get a fidelity of $0.946\pm0.002$, which is slightly lower than the value of original OAM-entangled state, due to the transferring process with OTPs. 

\begin{figure}[!htb]
	\centering
	\includegraphics[width=0.95\linewidth]{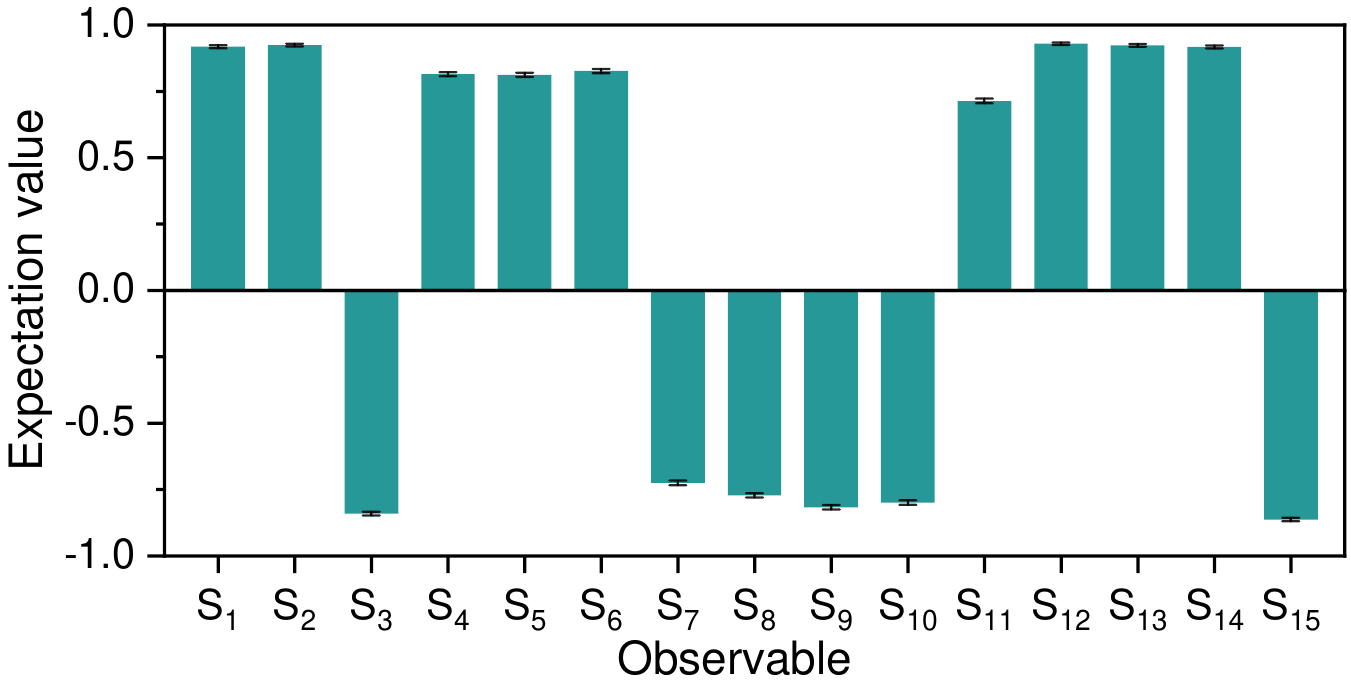}
	\caption{Experimental results for creating the hyperentangled photon pairs. The fidelity is calculated from the measured expectation values of 15 joint observables in two DOFs of polarization and OAM.}
	\label{fig5}
\end{figure}

In the above transformation processing, the entanglement will vanish in the DOF of OAM and appear in the DOF of polarization. Actually, our source can also support the polarization entanglement without eliminating the OAM entanglement, and produce the polarization-OAM hyperentangled photon pairs. Because the two down-converted photons have the same single frequency, we can construct the polarization entanglement by the postselection on a PBS as shown in Fig.~\ref{fig2}(b), which is the same as the previous construction process in Ref.~\cite{Bao2008}. Since this process does not destroy the OAM entanglement, as shown in Fig.~\ref{fig2}(c), the hyperentangled photon pairs can be generated in the form of  
\begin{equation}
\ket{\Psi}_{Hy}=\frac{1}{2}(\ket{H}_{A}\ket{H}_{B}+\ket{V}_{A}\ket{V}_{B})\otimes(\ket{+1}_{A}\ket{-1}_{B}+\ket{-1}_{A}\ket{+1}_{B}).
\label{eq7}
\end{equation}
Its density matrix $\rho_{Hy} = \ket{\Psi}_{Hy}\bra{\Psi}_{Hy}$ can be written as $\rho_{Hy} = \frac{1}{16}(II + X_pX_p - Y_pY_p + Z_pZ_p) \otimes (II + X_oX_o + Y_oY_o - Z_oZ_o)$, where $X_p$, $Y_p$ and $Z_p$ are the Pauli operators in the polarization DOF. With the measured expectation values of 15 joint observables \{$IIX_oX_o$, $IIY_oY_o$, $IIZ_oZ_o$, $X_pX_pII$, $X_pX_pX_oX_o$, $X_pX_pY_oY_o$, $X_pX_pZ_oZ_o$, $Y_pY_pII$, $Y_pY_pX_oX_o$, $Y_pY_pY_oY_o$, $Y_pY_pZ_oZ_o$, $Z_pZ_pII$, $Z_pZ_pX_oX_o$, $Z_pZ_pY_oY_o$, $Z_pZ_pZ_oZ_o$\} denoted as \{$S_1$, $S_2$, ..., $S_{15}$\}, as shown in Fig.~\ref{fig5}, we obtain a fidelity of $0.850\pm0.002$ for the generated hyperentangled state.

The experimental results shown in Figs.~\ref{fig3},~\ref{fig4} and~\ref{fig5} are obtained under the pump power of 50 mW. The corresponding brightness by the detected two-photon counts are 1400 s$^{-1}$ for the entanglement in the single DOF (OAM or polarization) and 700 s$^{-1}$ for the hyperentanglement due to the loss of 50\% in the postselection process. This pump power is much lower than the threshold of optical parametric oscillator~\cite{Boyd2020}, thus the generation rate of photon pairs is proportional to the pump power, which has been verified by the experimental results as shown in Fig.~S2 (see~\cite{SuppMat} for details). The best fitting gives the detected two-photon spectral brightness per pump power as 2.1 s$^{-1}$MHz$^{-1}$mW$^{-1}$, which can be greatly improved by double-pass~\cite{Chuu2012} or multi-pass (cavity resonance)~\cite{Mataji-Kojouri2023} pumping the nonlinear crystal, optimizing the pump condition, constructing a cavity with a higher finesse, and replacing the chopper with a more efficient way to eliminate the locking beam. The overall brightness of detected two-photon counts can be directly enhanced by increasing pump power at the cost of the fidelity of the entangled photon pairs owing to the increase of multi-pair emission. Therefore, the overall brightness above a fixed fidelity such as 0.9 is more important. From the visibility curve in Fig.~S2, we estimate that the fidelity of our OAM-entangled source will decrease to 0.9 when the pump power is increased to 100 mW, the corresponding detected two-photon brightness will be 2800 s$^{-1}$.

In summary, we have generated the first postselection-free narrow-band entangled photon pairs with cavity-enhanced SPDC in the dual-resonance cavity supporting the degenerate high-order OAM modes. The direct output OAM-entangled photon pairs have a linewidth of $\sim$13.8 MHz, a detected two-photon brightness of 1400 s$^{-1}$ and a fidelity of $0.969\pm0.003$. A competing technique to generate narrow-band entangled photon source with comparable apparatus complexity is SFWM in atom vapor~\cite{Chen2024}, which was also utilized to generate the narrow-band OAM-entangled photon source~\cite{Ma2024} very recently with spectral brightness of 16 s$^{-1}$MHz$^{-1}$ and a fidelity of 0.957. The spectral brightness of our narrow-band OAM-entangled photon source is 100 s$^{-1}$MHz$^{-1}$ under a similar fidelity of 0.969, showing the promising practical potential of our technique.
Due to the natural discrete high-dimensional feature of the OAM DOF, our scheme paves the way to produce the narrow-band high-dimensional entangled source~\cite{Erhard2020}, which is still an open challenge so far. Besides the direct application in quantum memory with cold atoms, our generated narrow-band entangled source opens new possibility for narrow-band-based quantum metrology~\cite{Giovannetti2011, Wolfgramm2013}. Our approach provides a convenient and effective route to produce the narrow-band entangled source at various wavelengths for wide applications such as solid-state quantum memory~\cite{Lago-Rivera2021, Liu2021, Jiang2023}.

\begin{acknowledgments}
This work was supported by the National Key R\&D Program of China (Grant No. 2020YFA0309500 and and No. 2019YFA0308700); the National Natural Science Foundation of China (Grants No. 12234009 and No. 12274215);  the Innovation Program for Quantum Science and Technology (Grant No. 2021ZD0301400); the Program for Innovative Talents and Entrepreneurs in Jiangsu; Key R\&D Program of Jiangsu Province (Grant No. BE2023002); the Key R\&D Program of Guangdong Province (Grant No. 2020B0303010001); the Natural Science Foundation of Jiangsu Province (Grant No. BK20220759).

\end{acknowledgments}



\begin{thebibliography}{10}


\bibitem{Chaneliere2005} T. Chaneli\`{e}re, D. N. Matsukevich, S. D. Jenkins, S.-Y. Lan, T. A. B. Kennedy, and A. Kuzmich, Storage and retrieval of single photons transmitted between remote quantum memories, Nature (London) \textbf{438}, 833 (2005).

\bibitem{Lvovsky2009} A. I. Lvovsky, B. C. Sanders, and W. Tittel, Optical quantum memory, Nat. Photonics  \textbf{3}, 706 (2009).

\bibitem{Duan2001} L. M. Duan, M. D. Lukin, J. I. Cirac, and P. Zoller, Long distance quantum communication with atomic ensembles
and linear optics, Nature (London) \textbf{414}, 413 (2001).

\bibitem{Sangouard2011} N. Sangouard, C. Simon, H. de Riedmatten, and N. Gisin, Quantum repeaters based on atomic ensembles and linear optics, Rev. Mod. Phys. \textbf{83}, 33 (2011).

\bibitem{Kok2007} P. Kok, W. J. Munro, K. Nemoto, T. C. Ralph, J. P. Dowling, and G. J. Milburn, Linear optical quantum computing with photonic qubits, Rev. Mod. Phys. \textbf{79}, 135 (2007).

\bibitem{Wehner2018} S. Wehner, D. Elkouss, and R. Hanson, Quantum internet: A vision for the road ahead, Science \textbf{362}, 303 (2018).

\bibitem{Azuma2023} K. Azuma, S. E. Economou, D. Elkouss, P. Hilaire, L. Jiang, H.-K. Lo, and I, Tzitrin, Quantum repeaters: From quantum networks to the quantum internet, Rev. Mod. Phys. \textbf{95}, 045006 (2023).

\bibitem{Zhang2011} H. Zhang et al., Preparation and Storage of Frequency-Uncorrelated Entangled Photons from Cavity-Enhanced Spontaneous Parametric Downconversion, Nat. Photonics  \textbf{5}, 628 (2011).

\bibitem{Ding2013} D. S. Ding, Z. Y. Zhou, B. S. Shi, and G. C. Guo, Single-photon-level quantum image memory based on cold atomic ensembles, Nat. Commun. \textbf{4}, 2527 (2013).

\bibitem{Nicolas2014} A. Nicolas, L. Veissier, L. Giner, E. Giacobino, D. Maxein, and J. Laurat, A Quantum Memory for Orbital Angular Momentum Photonic Qubits, Nat. Photonics \textbf{8}, 234 (2014).

\bibitem{Parigi2015} V. Parigi, V. D'Ambrosio, C. Arnold, L. Marrucci, F. Sciarrino, and J. Laurat, Storage and retrieval of vector beams of light in a multiple-degree-of-freedom quantum memory, Nat. Commun. \textbf{6}, 1 (2015).

\bibitem{Vernaz-Gris2018} P. Vernaz-Gris, K. Huang, M. Cao, A. S. Sheremet, and J. Laurat, Highly-efficient quantum memory for polarization qubits in a spatially-multiplexed cold atomic ensemble, Nat. Commun. \textbf{9}, 363 (2018).

\bibitem{Wang2019} Y. Wang, J. Li, S. Zhang, K. Su, Y. Zhou, K. Liao, S. Du, H. Yan, and S.-L. Zhu, Efficient quantum memory for single-photon polarization qubits, Nat. Photonics \textbf{13}, 346 (2019).

\bibitem{Ye2022} Y.-H. Ye, L. Zeng, M.-X. Dong, W.-H. Zhang, E.-Z. Li, D.-C. Li, G.-C. Guo, D.-S. Ding, and B.-S. Shi, Long-lived memory for orbital angular momentum quantum states, Phys. Rev. Lett. \textbf{129}, 193601 (2022).

\bibitem{Dong2023} M.-X. Dong, W.-H. Zhang, L. Zeng, Y.-H. Ye, D.-C. Li, G.-C. Guo, D.-S. Ding, B.-S. Shi, Highly Efficient Storage of 25-Dimensional Photonic Qudit in a Cold-Atom-Based Quantum Memory, Phys. Rev. Lett. \textbf{131}, 240801(2023).

\bibitem{Rakonjac2021} J. V. Rakonjac, D. Lago-Rivera, A. Seri, M. Mazzera, S. Grandi, and H. de Riedmatten, Entanglement between a Telecom Photon and an On-Demand Multimode Solid-State Quantum Memory, Phys. Rev. Lett. \textbf{127}, 210502 (2021).

\bibitem{Lago-Rivera2021} D. Lago-Rivera, S. Grandi, J. V. Rakonjac, A. Seri, and H. de Riedmatten, Telecom-heralded entanglement between multimode solid-state quantum memories, Nature (London) \textbf{594}, 37 (2021).

\bibitem{Rakonjac2022} J. V. Rakonjac, G. Corrielli, D. Lago-Rivera, A. Seri, M. Mazzera, S. Grandi, R. Osellame, and H. de Riedmatten, Storage and analysis of light-matter entanglement in a fiber-integrated system, Sci. Adv. \textbf{8}, 3919 (2022).

\bibitem{Ou1999}Z. Y. Ou and Y. J. Lu, Cavity Enhanced Spontaneous Parametric Down-Conversion for the Prolongation of Correlation Time between Conjugate Photons, Phys. Rev. Lett. \textbf{83}, 2556 (1999).

\bibitem{Scholz2009-OC} M. Scholz, L. Koch, and O. Benson, Analytical Treatment of Spectral Properties and Signal-Idler Intensity Correlations for a Double- Resonant Optical Parametric Oscillator Far below Threshold, Opt. Commun. \textbf{282}, 3518 (2009).

\bibitem{Slattery2019} O. Slattery, L. Ma, K. Zong, and X. Tang, Background and Review of Cavity-Enhanced Spontaneous Parametric Down-Conversion, J. Res. Natl. Ins. Stan. \textbf{124}, 124019 (2019).

\bibitem{Kuzmich2003} A. Kuzmich, W. P. Bowen, A. D. Boozer, A. Boca, C. W. Chou, L.-M. Duan, and H. J. Kimble, Generation of nonclassical photon pairs for scalable quantum communication with atomic ensembles, Nature (London) \textbf{423}, 731 (2003).

\bibitem{Thompson2006} J. K. Thompson, J. Simon, H. Loh, and V. Vuleti\'c, A high-brightness source of narrowband, identical-photon pairs, Science \textbf{313}, 74 (2006).


\bibitem{Yan2011} H. Yan, S. Zhang, J. F. Chen, M. M. T. Loy, G. K. L. Wong, and S. Du, Generation of Narrow-Band Hyperentangled Nondegenerate Paired Photons, Phys. Rev. Lett. \textbf{106}, 033601 (2011).

\bibitem{Srivathsan2013} B. Srivathsan, G. K. Gulati, B. Chng, G. Maslennikov, D. Matsukevich, and C. Kurtsiefer, Narrow Band Source of Transform-Limited Photon Pairs via Four-Wave Mixing in a Cold Atomic Ensemble, Phys. Rev. Lett. \textbf{111}, 123602 (2013).

\bibitem{Liao2014} K. Liao, Hui Yan, J. He, S. Du, Z.-M. Zhang, and S.-L. Zhu, Subnatural-Linewidth Polarization-Entangled Photon Pairs with Controllable Temporal Length, Phys. Rev. Lett. \textbf{112}, 243602 (2014). 

\bibitem{Chen2015} P. Chen, C. Shu, X. Guo, M. M. T. Loy, and S. Du, Measuring the Biphoton Temporal Wave Function with Polarization-Dependent and Time-Resolved Two-Photon Interference, Phys. Rev. Lett. \textbf{114}, 010401 (2015).

\bibitem{Lee2016} J.-C. Lee, K.-K. Park, T.-M. Zhao, and Y.-H. Kim, Einstein-Podolsky-Rosen Entanglement of Narrow-Band Photons from Cold Atoms, Phys. Rev. Lett. \textbf{117}, 250501 (2016).

\bibitem{Zhao2019}T.-M. Zhao, Y. S. Ihn, and Y.-H. Kim, Direct Generation of Narrow-Band Hyperentangled Photons, Phys. Rev. Lett. \textbf{122}, 123607 (2019).

\bibitem{Mei2020} Y. Mei, Y. Zhou, S. Zhang, J. Li, K. Liao, H. Yan , S.-L. Zhu, and S. Du, Einstein-Podolsky-Rosen Energy-Time Entanglement of Narrow-Band Biphotons, Phys. Rev. Lett. \textbf{124}, 010509 (2020).

\bibitem{Shu2016} C. Shu, P. Chen, T. K. A. Chow, L. Zhu, Y. Xiao, M. M. T. Loy, and S. Du, Subnatural-linewidth biphotons from a Dopplerbroadened hot atomic vapour cell, Nat. Commun. \textbf{7}, 12783 (2016).

\bibitem{Zhu2017}L. Zhu, X. Guo, C. Shu, H. Jeong, and S. Du, Bright narrowband biphoton generation from a hot rubidium atomic vapor cell, Appl. Phys. Lett. \textbf{110}, 161101 (2017).

\bibitem{WangC2018} C. Wang, Y. Gu, Y. Yu, D. Wei, P. Zhang, H. Gao, and F. Li, Efficient generation of non-classical photon pairs in a hot atomic ensemble, Chin. Opt. Lett. \textbf{16}, 082701 (2018).

\bibitem{Hsu2021} C.-Y. Hsu, Y.-S. Wang, J.-M. Chen, F.-C. Huang, Y.-T. Ke, E. K. Huang, W. Hung, K.-L. Chao, S.-S. Hsiao, Y.-H. Chen, C.-S. Chuu, Y.-C. Chen, Y.-F. Chen, and I. A. Yu, Generation of sub-MHz and spectrally-bright biphotons from hot atomic vapors with a phase mismatch-free scheme, Opt. Express \textbf{29}, 4632 (2021).

\bibitem{Chen2022} J.-M. Chen, C.-Y. Hsu, W.-K. Huang, S.-S. Hsiao, F.-C. Huang, Y.-H. Chen, C.-S. Chuu, Y.-C. Chen, Y.-F. Chen, and I. A. Yu, Room-temperature biphoton source with a spectral brightness near the ultimate limit, Phys. Rev. Res. \textbf{4}, 023132 (2022).

\bibitem{Ma2024} J. Ma, C. Wang, B. Li, Y. Chen, Y. Yang, J. Wang, X. Yang, S. Qiu, H. Gao, and F. Li, Generation of subnatural-linewidth orbital angular momentum entangled biphotons using a single driving laser in hot atoms, Opt. Express \textbf{32}, 23026 (2024).

\bibitem{Lu2003} Y. J. Lu, R. L. Campbell, and Z. Y. Ou, Mode-Locked Two-Photon States, Phys. Rev. Lett. \textbf{91}, 163602 (2003).

\bibitem{Liu2020} J. Liu, J. Liu, P. Yu, and G. Zhang, Sub-megahertz narrow-band photon pairs at 606 nm for solid-state quantum memories, APL Photonics \textbf{5}, 066105 (2020).

\bibitem{Kuklewicz2006} C. E. Kuklewicz, F. N. C. Wong, and J. H. Shapiro, Time-Bin-Modulated Biphotons from Cavity-Enhanced Down-Conversion, Phys. Rev. Lett. \textbf{97}, 223601 (2006).

\bibitem{Bao2008} X.-H. Bao, Y. Qian, J. Yang, H. Zhang, Z.-B. Chen, T. Yang, and J.-W. Pan, Generation of Narrow-Band Polarization-Entangled Photon Pairs for Atomic Quantum Memories, Phys. Rev. Lett. \textbf{101}, 190501 (2008).

\bibitem{Scholz2009} M. Scholz, L. Koch, and O. Benson, Statistics of Narrow-Band Single Photons for Quantum Memories Generated by Ultrabright Cavity-Enhanced Parametric Down-Conversion, Phys. Rev. Lett. 102, 063603 (2009).

\bibitem{Wolfgramm2011} F. Wolfgramm, Y. A. de Icaza Astiz, F. A. Beduini, A. Cer\`{e}, and M. W. Mitchell, Atom-Resonant Heralded Single Photons by Interaction-Free Measurement, Phys. Rev. Lett. \textbf{106}, 053602 (2011).

\bibitem{Rambach2016} M. Rambach, A. Nikolova, T. J. Weinhold, and A. G. White, Sub-megahertz linewidth single photon source, APL Photonics 1, 096101 (2016).

\bibitem{Moqanaki2019} A. Moqanaki, F. Massa, and P. Walther, Novel single-mode narrow-band photon source of high brightness tuned to cesium D2 line, APL Photonics \textbf{4}, 090804 (2019).

\bibitem{Prakash2021} V. Prakash, A. Sierant, and M. W. Mitchell, Autoheterodyne Characterization of Narrow-Band Photon Pairs, Phys. Rev. Lett. \textbf{127}, 043601 (2021).

\bibitem{Fekete2013} J. Fekete, D. Riel{\"a}nder, M. Cristiani, and H. de Riedmatten, Ultranarrow-Band Photon-Pair Source Compatible with Solid State Quantum Memories and Telecommunication Networks, Phys. Rev. Lett. 110, 220502 (2013).

\bibitem{Allen1992} L. Allen, M. W. Beijersbergen, R. J. C. Spreeuw, and J. P. Woerdman, "Orbital angular momentum of light and the transformation of Laguerre-Gaussian laser modes," Phys. Rev. A. \textbf{45}, 8185 (1992).

\bibitem{Hendrych2012} M. Hendrych, R. Gallego, M. Mi\v{c}uda, N. Brunner, A. A\'icn, and J. P. Torres, Experimental estimation of the dimension of classical and quantum systems, Nat. Phys. \textbf{8}, 588 (2012).

\bibitem{Malik2016} M. Malik, M. Erhard, M. Huber, M. Krenn, R. Fickler, and A. Zeilinger, Multi-photon entanglement in high dimensions, Nat. Photonics \textbf{10}, 248 (2016).

\bibitem{Bavaresco2018} J. Bavaresco, N. H. Valencia, C. Kl{\"o}ckl, M. Pivoluska, P. Erker, N. Friis, M. Malik, and M. Huber, Measurements in Two Bases Are Sufficient for Certifying High-Dimensional Entanglement, Nat. Phys. \textbf{14}, 1032 (2018).

\bibitem{Ecker2019} S. Ecker et al., Overcoming Noise in Entanglement Distribution, Phys. Rev. X \textbf{9}, 041042 (2019).

\bibitem{Mair2001} A. Mair, A. Vaziri, G. Weihs, and A. Zeilinger, Entanglement of the Orbital Angular Momentum States of Photons, Nature (London) \textbf{412}, (2001).

\bibitem{Pires2010} H. D. L. Pires, H. C. B. Florijn, and M. P. Van Exter, Measurement of the Spiral Spectrum of Entangled Two-Photon States, Phys. Rev. Lett. \textbf{104}, 020505 (2010).

\bibitem{Martinelli2004} M. Martinelli, J. A. O. Huguenin, P. Nussenzveig, and A. Z. Khoury, Orbital Angular Momentum Exchange in an Optical Parametric Oscillator, Phys. Rev. A \textbf{70}, 013812 (2004).

\bibitem{Navarrete-Benlloch2008} C. Navarrete-Benlloch, Eugenio Rold\'an, and Germa\'n J. de Valc\'arcel, Noncritically Squeezed Light via Spontaneous Rotational Symmetry Breaking, Phys. Rev. Lett. \textbf{100}, 203601 (2008).

\bibitem{Lassen2009} M. Lassen, G. Leuchs, and U. L. Andersen, Continuous Variable Entanglement and Squeezing of Orbital angular Momentum States, Phys. Rev. Lett. \textbf{102}, 163602 (2009).

\bibitem{Santos2009} B. C. dos Santos, K. Dechoum, and A. Z. Khoury, Continuous-Variable Hyperentanglement in A Parametric Oscillator with Orbital Angular Momentum, Phys. Rev. Lett. \textbf{103}, 230503 (2009).

\bibitem{Liu2014} K. Liu, J. Guo, C. Cai, S. Guo, and J. Gao, Experimental Generation of Continuous-Variable Hyperentanglement in an Optical Parametric Oscillator, Phys. Rev. Lett. \textbf{113}, 170501 (2014).

\bibitem{Cheng2018} Z.-D. Cheng, Q. Li, Z.-H. Liu, F.-F. Yan, S. Yu, J.-S. Tang, Z.-W. Zhou, J.-S. Xu, C.-F. Li, and G.-C. Guo, Experimental implementation of a degenerate optical resonator supportingmore than 46 Laguerre-Gaussian modes, Appl. Phys. Lett. \textbf{112}, 201104 (2018).

\bibitem{Zhou2017} Y. Zhou, M. Mirhosseini, D. Fu, J. Zhao, S. M. H. Rafsanjani, A. E. Willner, and R. W. Boyd, Sorting Photons by Radial Quantum Number, Phys. Rev. Lett. \textbf{119}, 263602 (2017).

\bibitem{Gu2018} X. Gu, M. Krenn, M. Erhard, and A. Zeilinger, Gouy Phase Radial Mode Sorter for Light: Concepts and Experiments, Phys. Rev. Lett. \textbf{120}, 103601 (2018).

\bibitem{Barros2021} R. F. Barros, G. B. Alves, and A. Z. Khoury, Gouy-Phase Effects in the Frequency Combs of an Optical Parametric Oscillator, Phys. Rev. A \textbf{103}, 023511 (2021).

\bibitem{Hiekkamlaki2022} M. Hiekkaml{\"a}ki, R. F. Barros, M. Ornigotti, and R. Fickler, Observation of the quantum Gouy phase, Nat. Photon. \textbf{16}, 828 (2022).

\bibitem{SuppMat} See Supplemental Material at http://link.aps.org/
supplemental/xx.yyyy/PhysRevLett.xxx.yyyy for Gouy phase and high-order OAM cavity, mode spectra of cavity resonant biphoton longitudinal modes, and brightness and visibility of detected narrow-band OAM-entangled photons vs the pump power.

\bibitem{Black2001} E. D. Black, An introduction to Pound-Drever-Hall laser frequency stabilization, Am. J. Phys. \textbf{69}, 79 (2001).

\bibitem{Wang2015} H. Wang, A. M. Marino, and J. Jing, Experimental implementation of phase locking in a nonlinear interferometer, Appl. Phys. Lett. \textbf{107}, 121106 (2015).

\bibitem{Boyd2020} R. W. Boyd, Nonlinear Optics, Fourth Edition (Academic Press, Inc., 2020).

\bibitem{Wang2015-N} X.-L. Wang, X.-D. Cai, Z.-E. Su, M.-C. Chen, D. Wu, L. Li, N.-L. Liu, C.-Y. Lu, and J.-W. Pan, Quantum Teleportation of Multiple Degrees of Freedom of a Single Photon, Nature(London) \textbf{518}, 516 (2015).

\bibitem{Wang2018} X.-L. Wang et al., 18-Qubit Entanglement with Six Photons' Three Degrees of Freedom, Phys. Rev. Lett. \textbf{120}, 260502 (2018).

\bibitem{Xu2022} J.-M. Xu, Y.-Z. Zhen, Y.-X. Yang, Z.-M. Cheng, Z.-C. Ren, K. Chen, X.-L. Wang, and H.-T. Wang, Experimental Demonstration of Quantum Pseudotelepathy, Phys. Rev. Lett. \textbf{129}, 050402 (2022).

\bibitem{James2001} D. F. V. James, P. G. Kwiat, W. J. Munro, and A. G. White, Measurement of qubits, Phys. Rev. A \textbf{64}, 052312 (2001).

\bibitem{Lou2024} Y.-C. Lou et al., Three-photon polarization entanglement of green light, Phys. Rev. Appl. \textbf{22}, 014052 (2024).

\bibitem{Chuu2012} C.-S. Chuu, G. Y. Yin, and S. E. Harris, miniature ultrabright source of temporally long, narrowband biphotons,  Appl. Phys. Lett. \textbf{101},  051108 (2012).

\bibitem{Mataji-Kojouri2023} A. Mataji-Kojouri and M. Liscidini, Narrow-band photon pair generation through cavity-enhanced spontaneous parametric down-conversion, Phys. Rev. A \textbf{108}, 053714 (2023).

\bibitem{Chen2024} J.-M. Chen, T. Peters, P.-H. Hsieh, and I. A. Yu, Review of biphoton source based on the double-$\Lambda$ spontaneous four-wave mixing process, Adv. Quantum Technol. \textbf{7}, 2400138 (2024).


\bibitem{Erhard2020} M. Erhard, M. Krenn, and A. Zeilinger, Advances in high-dimensional quantum entanglement, Nat. Rev. Phys. \textbf{2}, 365 (2020).

\bibitem{Giovannetti2011} V. Giovannetti, S. Lloyd, and L. Maccone, Advances in quantum metrology, Nat. Photonics \textbf{5}, 222 (2011).

\bibitem{Wolfgramm2013} F. Wolfgramm, C. Vitelli, F. A. Beduini, N. Godbout, and M. W. Mitchell, Entanglement-enhanced probing of a delicate material system, Nat. Photonics \textbf{7}, 28 (2013).

\bibitem{Liu2021} X. Liu, J. Hu, Z.-F. Li, X. Li, P.-Y. Li, P.-J. Liang, Z.-Q. Zhou, C.-F. Li, and G.-C. Guo, Heralded entanglement distribution between two absorptive quantum memories, Nature (London) \textbf{594}, 41 (2021).

\bibitem{Jiang2023} M.-H. Jiang, W. Xue, Q. He, Y.-Y. An, X. Zheng, W.-J. Xu, Y.-B. Xie, Y. Lu, S. Zhu, and X.-S. Ma, Quantum storage of entangled photons at telecom wavelengths in a crystal, Nat. Commun. \textbf{14}, 6995 (2023).

\end{thebibliography}
\end{document}